\documentclass[journal = nalefd, 
manuscript = letter, 
layout = traditional, 
email = true,
super = false,
articletitle = true,
chaptertitle = true,
]{achemso}

\setkeys{acs}{doi = true}

\usepackage{threeparttable}
\usepackage{setspace}
\usepackage{makecell}
\usepackage{amsmath}
\usepackage{graphicx}
\usepackage{dcolumn}
\usepackage{bm}
\usepackage{hyperref}
\usepackage[mathlines]{lineno}
\usepackage{textcomp}
\usepackage{gensymb}
\usepackage{subfigure}
\usepackage{url}
\usepackage{xcolor}
\usepackage{hhline}
\usepackage[version=3]{mhchem}
\usepackage{multirow}
\usepackage[american]{babel}
\usepackage{hhline}
\usepackage{textgreek}
\usepackage{miller}
\usepackage{lipsum}

\definecolor{red}{rgb}{0.6,0.1,0.1}
\definecolor{blue}{rgb}{0.0,0.0,0.0}\def\blue{\color{blue}}
\definecolor{green}{rgb}{0.1,0.6,0.1}
\definecolor{yellow}{rgb}{0.8,0.8,0.6}

\author{Qi Li$^{\nabla}$} 
\affiliation{State Key Laboratory of Crystal Materials, Institute of Novel Semiconductors, Institute of Crystal materials, Shandong University, Jinan, Shandong, 250100, China}

\author{Junlei Zhao$^{\nabla}$}
\email{zhaojl@sustech.edu.cn}
\affiliation{Department of Electronic and Electrical Engineering, Southern University of Science and Technology, Shenzhen, 518055, China}

\author{Na Lin}
\email{linna@sdu.edu.cn}
\affiliation{State Key Laboratory of Crystal Materials, Institute of Novel Semiconductors, Institute of Crystal materials, Shandong University, Jinan, Shandong, 250100, China}

\author{Xiufeng Cheng}
\affiliation{State Key Laboratory of Crystal Materials, Institute of Novel Semiconductors, Institute of Crystal materials, Shandong University, Jinan, Shandong, 250100, China}

\author{Xian Zhao} 
\affiliation{State Key Laboratory of Crystal Materials, Institute of Novel Semiconductors, Institute of Crystal materials, Shandong University, Jinan, Shandong, 250100, China}
\alsoaffiliation{Center for Optics Research and Engineering, Shandong University, Qingdao, Shandong, 266237, China}

\author{Zhaojun Liu} 
\affiliation{Department of Electronic and Electrical Engineering, Southern University of Science and Technology, Shenzhen, 518055, China}

\author{Zhitai Jia}
\email{z.jia@sdu.edu.cn}
\affiliation{State Key Laboratory of Crystal Materials, Institute of Novel Semiconductors, Institute of Crystal materials, Shandong University, Jinan, Shandong, 250100, China}
\alsoaffiliation{Shandong Research Institute of Industrial Technology, Jinan, Shandong, 250100, China}

\author{Mengyuan Hua} 
\email{huamy@sustech.edu.cn}
\affiliation{Department of Electronic and Electrical Engineering, Southern University of Science and Technology, Shenzhen, 518055, China}

\title{{\blue Edge-Dependent Step-Flow Growth Mechanism in \texorpdfstring{\textit{\textbeta}-\ce{Ga2O3}}{} (100) Facet at the Atomic Level}}


\begin{document}



\begin{abstract}

{\blue Homoepitaxial step-flow growth of high-quality $\beta$-\ce{Ga2O3} thin films is essential for the advancement of high-performance \ce{Ga2O3}-based devices.}
In this work, the step-flow growth mechanism of $\beta$-\ce{Ga2O3} \hkl(100) facet is explored by machine-learning molecular dynamics simulations and density functional theory calculations. 
{\blue Our results reveal that Ga adatoms and Ga-O adatom pairs, with their high mobility, are the primary atomic species responsible for efficient surface migration on the \hkl(100) facet. 
The asymmetric monoclinic structure of $\beta$-\ce{Ga2O3} induces a distinct two-stage Ehrlich-Schwoebel barrier for Ga adatoms at the \hkl[00-1] step edge, contributing to the suppression of double-step and hillock formation.
Furthermore, a miscut towards \hkl[00-1] does not induce the nucleation of stable twin boundaries, whereas a miscut towards \hkl[001] leads to the spontaneous formation of twin boundaries.} 
This research provides meaningful insights not only for high-quality $\beta$-\ce{Ga2O3} homoepitaxy but also the step-flow growth mechanism of other similar systems.

\end{abstract}

\maketitle

{\blue Gallium oxide (\ce{Ga2O3}) has garnered enormous interest from (opto-)electronics  community as a promising next-generation ultrawide bandgap semiconductor.}
Among its polymorphs, the monoclinic $\beta$-\ce{Ga2O3}, the most stable phase under ambient conditions, exhibits several remarkable properties, including an ultrawide bandgap ($E_\mathrm{g} \simeq 4.9$~eV), a high breakdown electric field ($E_\mathrm{crit} \simeq 8$~MV/cm), exceptional chemical and thermal stability (melting point $\simeq 1720$~\degree C), a large Baliga's figure of merit ($\mathrm{BFOM} = 3444$), and a short ultraviolet cutoff edge of $\sim 260$~nm~\cite{apl2008villora, jsap2015onuma, pearton2018a, acspho2018arora, aom2021kaur}. 
Owing to these outstanding material properties, $\beta$-\ce{Ga2O3} possesses great prospects for applications in high-power electronics~\cite{jcg2013kohei, nc2022zhang}, solar-blind ultraviolet optoelectronics~\cite{apl2018higashiwaki, acsnano2021wang}, {\blue and two-dimensional devices~\cite{aljarb2020ledge, SFzhao2021two, yi2024integration}}. 
To fabricate high-performance $\beta$-\ce{Ga2O3}-based device, a high-quality epitaxial thin film with a smooth surface and low defect density is essential~\cite{jap2019vaidya, apl2020Bin, aplm2020mauze}. 
Particularly, homoepitaxy is significant for minimizing defect density in the $\beta$-\ce{Ga2O3} epitaxial thin film by avoiding lattice mismatch in heteroepitaxy. 
Recently, the homoepitaxy of $\beta$-\ce{Ga2O3} has been explored using several thin film growth techniques, such as metal-organic vapor phase epitaxy (MOVPE)~\cite{apl2020Bin, jap2023chouta} (also referred to as metal-organic chemical vapor deposition, MOCVD, in the literature~\cite{sm2022tao, cgd2024meng}), molecular beam epitaxy (MBE)~\cite{aplm2020mauze, mtp2024kuang}, halide vapor phase epitaxy (HVPE)~\cite{aplm2018leach, apl2021sayleap}, and mist chemical vapor deposition (mist-CVD)~\cite{jjap2020isomura, mssp2021hiroyuki}. 

Extensive literature has demonstrated that the facet orientation of $\beta$-\ce{Ga2O3} homoepitaxial substrate can significantly affect the growth kinetics, and hence the quality of the homoepitaxial thin film~\cite{apl2019feng, aplm2020mazzolini, jac2020trong, jvsta2022Meng, apl2022ken, apl2024zhang}. 
Among the commonly studied facet orientations, namely, \hkl(100), \hkl(010), \hkl(001), and \hkl(-201), the \hkl(100) facet is the easiest to prepare and is considered as the most adequate orientation for the devices requiring smooth surfaces and/or high-quality interfaces. 
This is because the \hkl(100) plane serves as the preferred cleavage plane with the lowest surface energy in $\beta$-\ce{Ga2O3}~\cite{apl2020musai, aipad2021chouta}. 
However, the main challenge associated with homoepitaxy on the \hkl(100) $\beta$-\ce{Ga2O3} substrate is the possibility of forming twin boundary with high density due to \ce{Ga}-sublattice double positioning~\cite{jap2016schewski, pccp2024qi}. 
As a remedy, it has been demonstrated that by introducing miscuts towards \hkl[00-1] on substrates oriented to \hkl(100) plane, the epitaxial growth of $\beta$-\ce{Ga2O3} can achieve stable step-flow growth and obtain epitaxial thin films with excellent crystal quality~\cite{jap2016schewski, aplm2019schewski, apl2020Bin, apl2020piero, jap2023chouta, chou2025impurity}. 

{\blue The step-flow growth mode is an epitaxial process in which deposited adatoms migrate to step edges between adjacent atomic layers on a surface before nucleating on terraces. 
This mechanism enables smooth, layer-by-layer growth while preventing the formation of hillocks on terraces, ultimately resulting in smooth film surfaces with high crystalline quality~\cite{mser1997hiroyuki, prl2005hong}.
Therefore, exploring the atomic mechanisms underlying the step-flow growth and achieving morphology control in $\beta$-\ce{Ga2O3} homoepitaxy is of great significance from both fundamental and application perspectives. 
A particularly important factor in step-flow epitaxial growth is the Erhlich-Schwoebel (ES) barrier (denoted as $E_{\mathrm{ES}}$), which refers to the additional energy barrier for adatoms to overcome when diffusing across a surface step edge~\cite{jap1966schwoebel, jcp1966ehrlich, jap1969schwoebel}. 
A large $E_{\mathrm{ES}}$ can lead to step-edge meandering and terrace morphology alteration, potentially destabilizing the step-flow growth~\cite{jpcm1994pimpine, prl2001paulin, jcg2012magdalena}.} 

In recent studies, numerical models~\cite{jap2016schewski} have been developed and experimentally validated to describe the morphology control of \hkl(100) $\beta$-\ce{Ga2O3} films grown by MOVPE~\cite{apl2020Bin, jap2023chouta}. 
However, the exploration of the complex kinetic and thermodynamic evolution of the step-flow growth processes at the atomic scale is still lacking. 
Furthermore, the migration barrier and path of adatoms on the surface of $\beta$-\ce{Ga2O3} substrates must be clarified to determine the $E_{\mathrm{ES}}$ and explain the essential mechanism of the step-flow growth of $\beta$-\ce{Ga2O3}.
In this work, the mechanism of the step-flow growth of $\beta$-\ce{Ga2O3} on the \hkl(100) substrates with different miscuts was demonstrated at the atomic resolution. 
Specifically, the minimum energy paths (MEPs) and minimum energy barriers (MEBs) of adatoms on the \hkl(100) $\beta$-\ce{Ga2O3} substrates with and without miscuts were revealed using nudged elastic band (NEB) method.
The complex dynamic evolution and accurate energetics were explored using machine-learning molecular dynamics (ML-MD) simulations, and density functional theory (DFT) calculations.
{\blue Furthermore, we emphasized that the inherently asymmetric atomic configurations at the miscut \hkl[00-1] and \hkl[001] step edges result in distinct kinetic and thermodynamic evolution of Ga and O atomic migrations. 
Additionally, the differences in nucleation energies of twin boundaries (TBs) at these two different miscut step edges provide a compelling explanation for the variations in epitaxy quality observed in recent experimental studies~\cite{aplm2019schewski, apl2020Bin, jap2023chouta}.} 

\begin{figure*}[htbp!]
    \includegraphics[width=16cm]{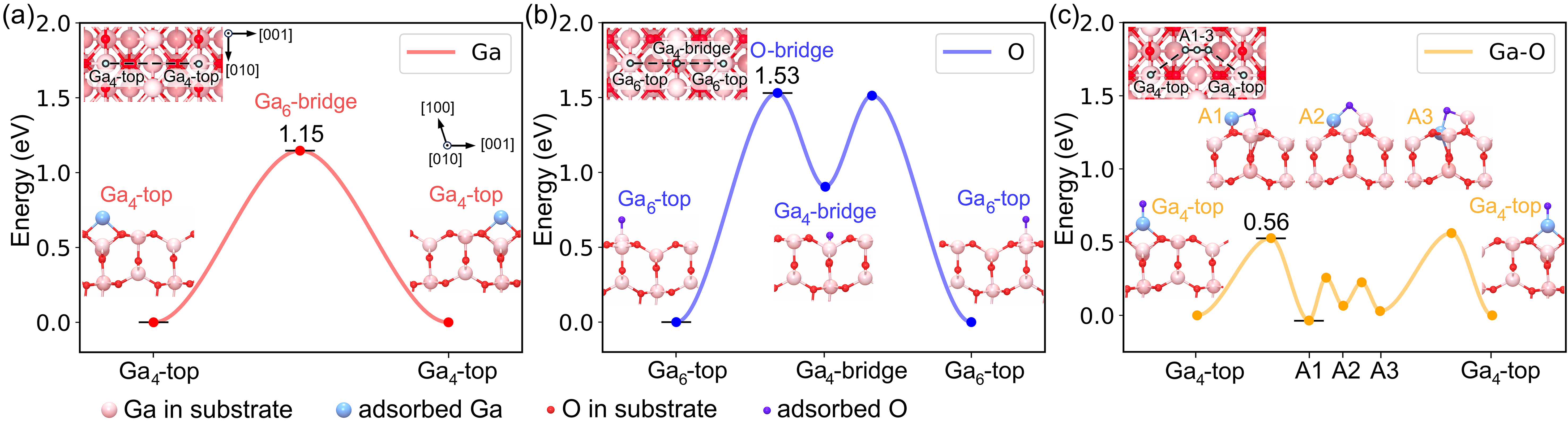}
    \caption{
    {\blue MEBs for (a) a Ga adatom (big lightblue sphere), (b) an O adatom (small purple sphere), and (c) a Ga-O adatom pair on the \hkl(100) surface of $\beta$-\ce{Ga2O3}.
    Substrate Ga and O atoms are shown in pink and red, respectively.
    Insets at the top-left corners of the three panels provide the top views of the corresponding MEPs. 
    The curves represent the fitted MEBs, with each valley point corresponding to the adjacent structure.
    Ga$_{4}$ and Ga$_{6}$ denote the four-coordinated and six-coordinated Ga substrate sites, respectively.}
    }
    \label{fig:1}
   \end{figure*}

{\blue Before investigating atomic migrations at the miscut step edges, we first examine the migration behavior of Ga/O adatoms and Ga-O adatom pair on the flat \hkl(100) surface of $\beta$-\ce{Ga2O3}, as predicted by the ML model, as illustrated in Figure~\ref{fig:1}. 
We conducted a comprehensive search of atomic migration paths (detailed in Supporting Information Note~1, Figures~S1 and S2, and Table~S1), closely examining all the possible direct hopping/exchange migration paths. 
The MEPs in the \hkl[001]/\hkl[00-1] directions on the \hkl(100) surface involve direct hopping for the Ga/O adatoms (Figure~\ref{fig:1}a and \ref{fig:1}b) and Ga exchange for the Ga-O adatom pair (Figure~\ref{fig:1}c).}
The corresponding MEBs for Ga adatom, O adatom, and Ga-O adatom pair were found to be 1.15 eV, 1.53 eV, and 0.56 eV, respectively.
{\blue We also validated these ML-predicted MEPs and MEBs through DFT calculations (Supporting Information Note~1, Figure~S4), which exhibited the same trends with reasonable deviation.} 
Based on the ML and DFT results, the mobility of Ga adatom is consistently higher than that of O adatom.
The Ga-O adatom pair can enhance the migration of surface adatoms, by facilitating the exchange migration path. 
These results provide important references for the subsequent exploration of atomic migration on the \hkl(100) surface with different miscut step edges.

\begin{figure*}[ht!]
    \includegraphics[width=16cm]{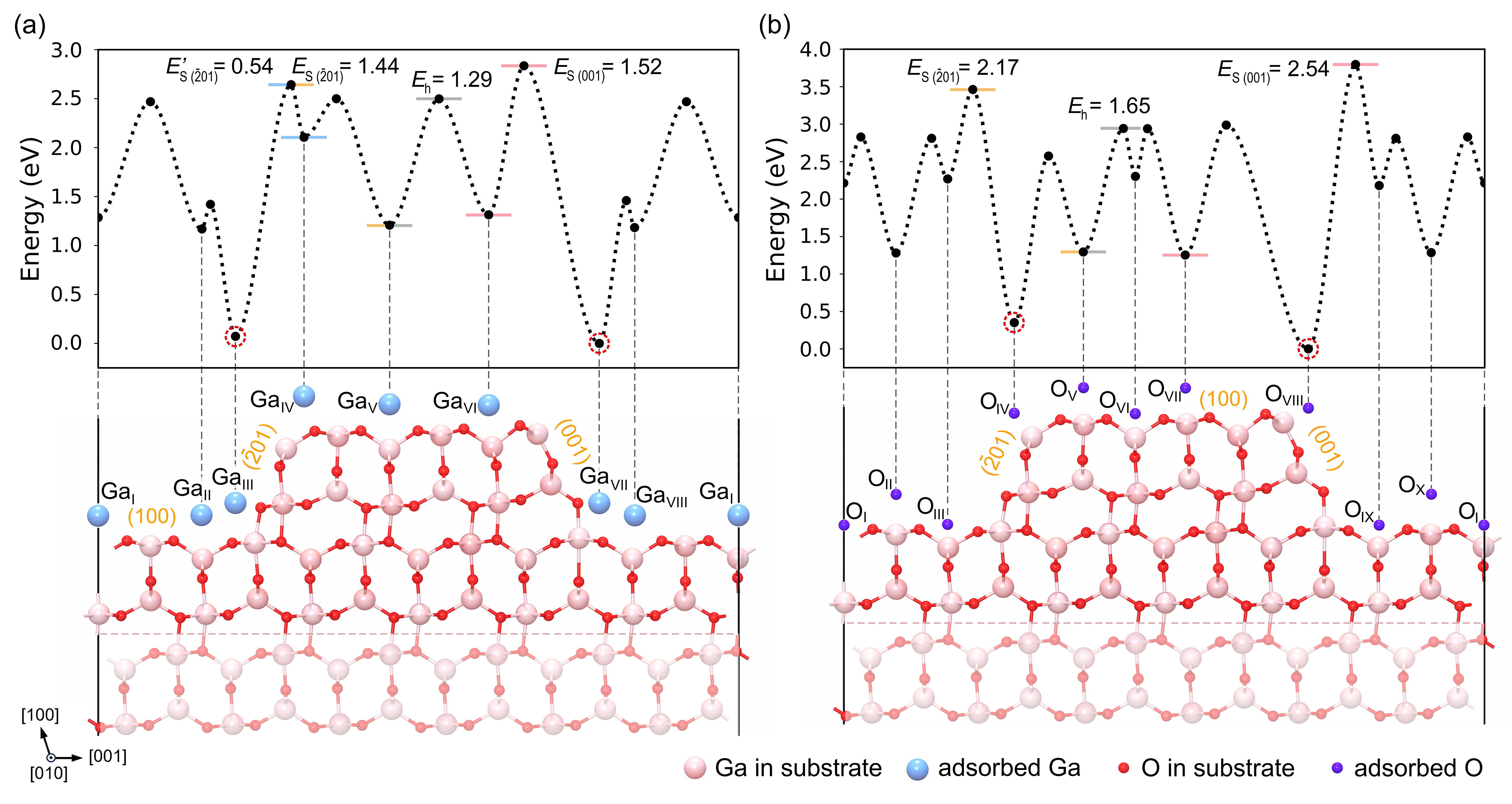}
    \caption{
    The overall MEBs of (a) Ga and (b) O adatoms on the \hkl(100) surface with two miscut step edges.
    The black dashed curves in the upper panels represent the fitted MEBs. 
    The valley points of the curves correspond to metastable adsorption sites which are represented as blue or purple spheres for adsorbed Ga or O, respectively. 
    The lowest energy sites at the two step edges are marked with red dashed circles.
    The zero energy is leveled to the global energy minima of the two curves.
   }
    \label{fig:2}
   \end{figure*}

We further investigated the MEPs and MEBs of Ga/O adatom migration on the $\beta$-\ce{Ga2O3} \hkl(100) surface with two miscut step edges, as shown in Figure~\ref{fig:2}.
The left step is a \hkl[00-1] step with a \hkl(-201) step-edge facet, and the right step is a \hkl[001] step with a \hkl(001) step-edge facet. 
The atomic migration paths at the step edges on the both sides were searched systematically (see Supporting Information, Note 2, Figure~S5, and Tables~S2 and S3).
The MEPs on the terrace are the same as those on the flat \hkl(100) surface (Figure~\ref{fig:1}), whereas the MEBs exhibit marginal increases (1.29 eV \textit{vs.} 1.15 eV for Ga and 1.65 eV \textit{vs.} 1.53 eV for O), owing to the compressive strain effect of the stepped terrace. 

{\blue Importantly, the most striking observation is the distinct single- and two-stage ES barriers of Ga adatom, rooted from the asymmetric monoclinic lattice of $\beta$-\ce{Ga2O3}. 
As shown in Figure~\ref{fig:2}a, the overall MEB for migrating downward the \hkl[00-1] step edge ($\mathrm{Ga_{V}} \rightarrow \mathrm{Ga_{IV}} \rightarrow \mathrm{Ga_{III}}$) is $E_{\mathrm{S\:\hkl(-201)}} = 1.44$~eV. 
In comparison to the MEB climbing downward the \hkl[001] step edge ($\mathrm{Ga_{VI}} \rightarrow \mathrm{Ga_{VII}}$, $E_{\mathrm{S\:\hkl(001)}}=1.52$~eV), the energy difference in the ES barriers is not significant (0.15 eV \textit{vs.} 0.23 eV).
However, a two-stage MEP is revealed for migrating downward the \hkl[00-1] step edge, where the $\mathrm{Ga_{IV}}$ site serves as a metastable shallow valley with an effective downhill barrier of $E'_{\mathrm{S\:\hkl(-201)}}=0.54$~eV. 
In addition to the overall smaller \hkl[00-1] ES barrier (0.15~eV), this intermediate $\mathrm{Ga_{IV}}$ site (similar to an intermediate state in catalysis) can significantly promote the downhill migration of Ga adatom at the \hkl[00-1] step edge, thereby suppressing the formation of double-steps and hillocks on the terrace in the \hkl[00-1] miscut direction.} 

{\blue Additionally, the energies of Ga adatoms at the step-edge-bottom sites ($\mathrm{Ga_{III}}$ and $\mathrm{Ga_{VII}}$) are much lower, compared to the energies at the step-edge-top sites ($\mathrm{Ga_{IV}}$ and $\mathrm{Ga_{VI}}$). 
These energy differences lead to a strong downhill migration trend for Ga adatoms at the step edge. 
These results have been validated by DFT calculations, which can be found in Supporting Information, Note~2, Figures~S5 and S7. 
For comparison, we also calculated the MEBs of Ga adatoms on a step with the \hkl(-201) terrace and \hkl(100) step-edge surface (see Supporting Information, Note~2, Figure~S8), showing a much higher MEBs and ES barriers. 
On the other hand, the MEBs of O adatom migration on both terrace and step edges are higher than those of Ga adtom, as shown in Figure~\ref{fig:2}b.
Notably, unlike the stable Ga sites, the most stable sites near the step edges for O adatoms are the step-edge-top sites, which is expected for anions, as the step-edge-bottom sites are screened by the O anions carrying the same type of charge.}  

However, although the NEB calculations for the migration of Ga and O adatoms provide clear evidence on the edge-dependent step-flow growth mechanism, arising from the asymmetry of $\beta$-\ce{Ga2O3} lattice, the manybody (such as Ga-O adatom pair and more) migration is too complicated to be effectively described by MEPs and MEBs.
Therefore, we further conducted ML-MD simulations to elucidate the dynamic migration process for Ga/O adatoms, and Ga-O adatom pairs on a large-sized $\beta$-\ce{Ga2O3} \hkl(100) facet with a miscut \hkl[00-1] step edge. 

\begin{figure*}[ht!]
    \includegraphics[width=16cm]{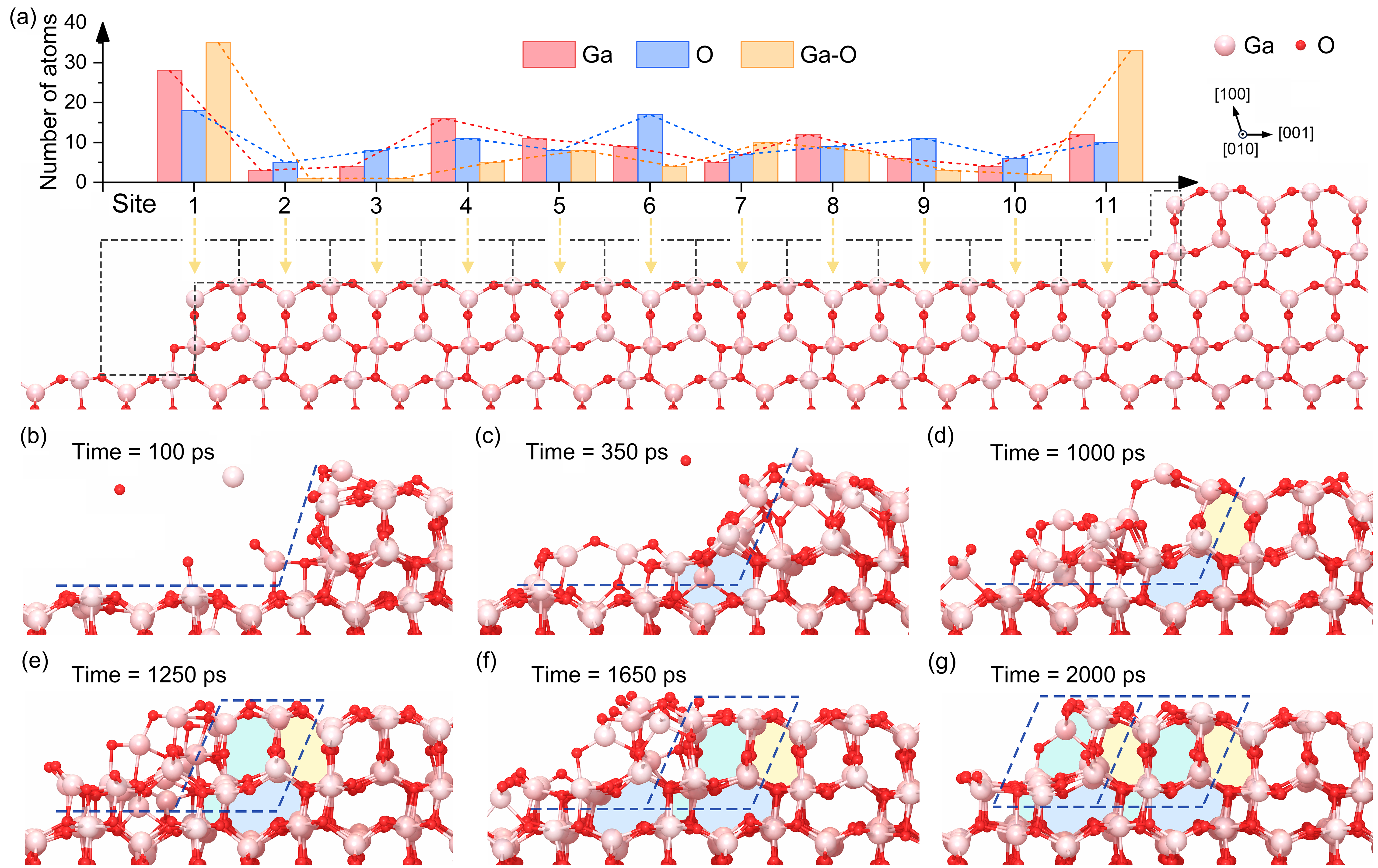}
    \caption{
    {\blue The ML-MD simulations of atomic migration and epitaxial step-flow growth on the \hkl(100) substrate with a \hkl[00-1] miscut step. 
    (a) The number of Ga/O adatoms and Ga-O adatom pairs migrating to different site regions after annealing for 1500 ps at 1200 K. 
    Each type of adatom (and pair) is deposited 10 times at all the 11 sites marked by the yellow arrows. 
    (b-g) The snapshots of step-flow growth of $\beta$-\ce{Ga2O3} on the \hkl(100)-\hkl[00-1] step at 1200 K. 
    The dashed lines in panels (b-g) distinguish the initial substrate.
    The colored shadows indicate the atomic rings characteristic of $\beta$-\ce{Ga2O3}.
    A completed periodic unit of epitaxial \hkl(-201) step edge is highlighted by a blue dashed parallelogram.
    See Supporting Information, Movie S1, for the full process.}
   }
    \label{fig:3}
\end{figure*}

{\blue As shown in Figure~\ref{fig:3}a, in the random adatom migration simulations, Ga/O adatoms and Ga-O adatom pairs were deposited 10 times at each of the 11 different deposition sites, marked by yellow arrows.
The final positions of the adatoms (and pairs) were tracked and counted after annealing for 1500~ps at 1200~K (see Supporting Information, Note~3 for details).
The Ga-O adatom pairs tend to migrate quickly to the more stable vicinal sites of the step edges (Sites 1 and 11), exhibiting the highest mobility among the three species tested. 
Ga adatoms show the second-highest migration mobility, as indicated by the higher count at Site 1, while the distribution of the final O adatom sites remains relatively uniform, with no significant atomic migration.
These statistics are in good agreement with the MEBs shown in Figure~\ref{fig:1}.}
{\blue Notably, the relatively high mobility of Ga adatom and Ga-O adatom pair leads to an essential assumption for the subsequent deposition ML-MD simulations, which consider terrace widths of a few nanometers and time spans of a few nanoseconds: Within a reasonable MD simulation time, the adatoms can migrate to the step-edge region across the full width of the terrace. 
Consequently, the epitaxial step-flow growth is insensitive to the initial deposition sites of adatoms, as low-mobility O adatoms can be mobilized by forming Ga-O adatom pairs upon encountering Ga adatoms.} 

{\blue Based on this observation and discussion, we further simulated the dynamic processes of epitaxial growth at the $\beta$-\ce{Ga2O3} \hkl(100)-\hkl[00-1] step edges, as shown in the snapshots in Figure~\ref{fig:3}b-g (More details in Supporting Information, Note 3, Figures~S9 and S10, and Movies~S1 and S2). 
The deposited adatoms near the step edge quickly bind to the edge-bottom site, and then grow epitaxially, completing the atomic rings in the sequence indicated by the ``blue $\rightarrow$ yellow $\rightarrow$ cyan $\rightarrow$ blue'' shadows in Figure~\ref{fig:3}b-g. 
The epitaxial layer initially grows a half-layer extending the step edge (blue shadow in Figure~\ref{fig:3}c) before forming the upper half-layer near the step-edge surface (yellow and cyan shadows in Figure~\ref{fig:3}d and \ref{fig:3}e). 
Once one period of the epitaxial \hkl(-201) step edge is completed, a second \hkl(-201) step-edge unit can be formed, repeating the step-flow growth process in the \hkl[00-1] direction (depicted by the additional blue, yellow, and cyan shadows in Figure~\ref{fig:3}f and \ref{fig:3}g). }

Meanwhile, the step-flow growth process at the \hkl(100)-\hkl[001] step edge can be investigated by ML-MD simulation, as shown in Supporting Information, Note 3, Figure~S10. 
Notably, in contrast to the perfect step-flow growth observed at the \hkl[00-1] step edge, the nucleation of TBs was observed in the simulations at the \hkl[001] step edge. 
These findings are consistent with the previous experimental observations~\cite{jap2016schewski, aplm2019schewski}, which reported significant lower quality of $\beta$-\ce{Ga2O3} epitaxial films grown on the \hkl(100)-\hkl[001] step substrates compared to those grown on the \hkl(100)-\hkl[00-1] step substrates, primarily due to the high density of TBs. 
Our ML-MD simulations suggest that these TBs in the epitaxial films may originate from the deposited atoms near the \hkl[001] step edge or even from the structural transformations of the step edge itself.

\begin{figure*}[ht!]
    \includegraphics[width=16cm]{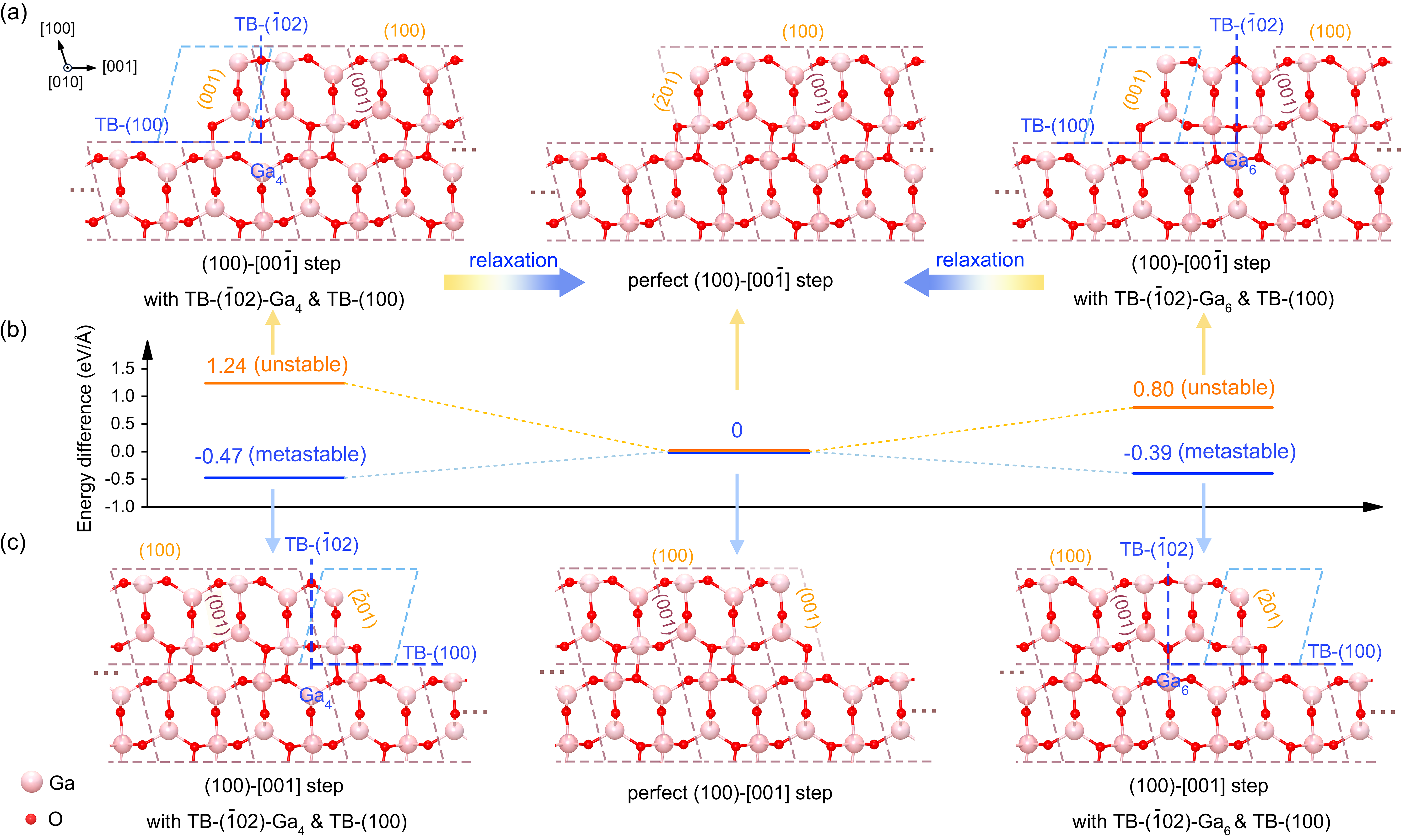}
    \caption{
    The TB structures of $\beta$-\ce{Ga2O3} at (a) \hkl(100)-\hkl[00-1] step edges and (c) \hkl(100)-\hkl[001] step edges, and (b) the energy differences between these TB structures and the perfect step.
    The TBs are represented by dark blue dashed lines. 
    The perfect lattice and twins lattice are marked with pink and light blue dashed parallelogram boxes, respectively.
   }
    \label{fig:4}
\end{figure*}

To explain why it is easier to form TBs on the \hkl[001] step compared to the \hkl[00-1] step, we conducted further static energy calculations for TB nucleation using the ML model. 
DFT calculations were also employed to ensure higher accuracy and provide cross-validation.  
As shown in Figure~\ref{fig:4}, we constructed the TBs on both the \hkl[00-1] and \hkl[001] step edges, considering two different types of atomic structures of TBs at each step edge. 
The two different TB structures for each step edge are distinguished by the different positions of TB-\hkl(-102) (Ga$_{4}$ and Ga$_{6}$ sites), as each twin at the step edge is composed of TB-\hkl(-102) and TB-\hkl(100). 
The relaxed structures of these steps with TBs are shown in Supporting Information, Note~4, Figure~S11. 
The energy differences between the TB and perfect step configurations are shown in Figure~\ref{fig:4}b. 
For the \hkl[00-1] step, both TB structures shown in Figure~\ref{fig:4}a are energetically unstable and spontaneously relax back into the perfect step after structure relaxation. 
On the contrary, for the \hkl[001] step, both TB structures shown in Figure~\ref{fig:4}c are, rather surprisingly, energetically favorable, with energy differences of $-0.47$~eV/Å and $-0.39$~eV/Å compared to the perfect step. 
Therefore, while TBs are energetically unfavorable to form on the \hkl[00-1] step edge, they are energetically favorable and easy to form on the \hkl[001] step edge. 
Our results provide a direct explanation for the phenomena observed in previous experiments~\cite{jap2016schewski, aplm2019schewski} as well as in our ML-MD simulations. 

Therefore, to achieve the high quality step-flow growth in the homoepitaxy of $\beta$-\ce{Ga2O3}, the selection of the orientation of substrates and the directions of miscuts are essential, which determines the terrace facet and step edge. 
It is important that atoms can migrate to the step-edge site efficiently, which requires that the migration energy of atoms on the terrace surface is low and the energy of atoms at the step-edge site is significantly lower than that of other sites, as shown in Figure~\ref{fig:2} and Figure~\ref{fig:3}a. 
Furthermore, both the terrace and step edge surfaces should have good thermal stability to maintain the regular step surface. 
Notably, even when the above requirements are satisfied, an improper miscut direction can promote the formation of TBs. 
For instance, the \hkl(100)-\hkl[001] step edges illustrated in Figure~\ref{fig:4} can facilitate TBs formation. 
In contrast, the atomic structures of the \hkl(100)-\hkl[00-1] step edges could effectively prevent the occurrence of TBs. 

In this work, we investigated the atomic mechanisms of the step-flow growth on $\beta$-\ce{Ga2O3} \hkl(100) facet using ML-MD and DFT methods. 
Ga adatom and Ga-O adatom pair are the primary atomic species responsible for efficient atomic migration on the \hkl(100) surface, both with and without a miscut, owing to their relatively low MEBs. 
The intrinsic lattice asymmetry of $\beta$-\ce{Ga2O3} leads to a distinct two-stage ES barrier of Ga adatoms at the \hkl[00-1] step edge, favoring downhill migration towards the stable step-edge-bottom site. 
The periodic step-flow growth processes on the \hkl(100)-\hkl[00-1] and \hkl(100)-\hkl[001] step edges are revealed through ML-MD simulations at atomic resolution. 
Furthermore, the TB nucleation structures at the \hkl(100)-\hkl[001] step edge are proved to be energetically more favorable than the perfect step, while the TB structures on the \hkl(100)-\hkl[00-1] step edges are unstable. 
Our study clarifies the mechanisms of the step-flow growth on $\beta$-\ce{Ga2O3} \hkl(100) substrates, providing an atom-level theoretical basis for the selection and processing of substrates to achieve high-quality $\beta$-\ce{Ga2O3} epitaxial
thin films.

\section{Methodology} 

The ML-MD calculations in this word are performed using the Large-scale Atomic/Molecular Massively Parallel Simulator (LAMMPS) code~\cite{lammps2022}. 
The ML interatomic potential was recently developed in our previous work~\cite{npjcm2023zhao} makes the large-scale MD simulations of \ce{Ga2O3} homoepitaxy feasible~\cite{apl2024zhang} and high-energy collision cascade. 
The atomic migration MEBs were calculated by the climbing-image nudged elastic band (CI-NEB) method with improved tangent estimation~\cite{jcp2000Henkelman, henkelman2000improved}, converging the forces to within $1 \times 10^{-3}$~eV/\r A and energies to within $1 \times 10^{-6}$~eV. 
These NEB calculations were performed using 9 images between the initial and final structures. 
For MD simulations of the $\beta$-\ce{Ga2O3} homoepitaxy and the migration of Ga/O atoms or Ga-O atomic pair on the step surface, the temperature were set as 1200~K. 
All the MD simulation cells consist of three groups of atoms depending on the initial positions: (i) the atoms in the fixed layer are fixed to avoid movement of the entire cell due to the momentum introduced by the incoming deposited atoms; (ii) the atoms in the thermal layer are controlled by Nos{\'e}-Hoover thermostat~\cite{Nose10061984, PhysRevA.31.1695}; and (iii) the atoms in the surface layer, epilayer, and deposition zone are allowed to move freely following Newton's law.
More details on ML-MD calculations can be found in the Supporting Information, Note~3.

All of our DFT calculations are performed using Vienna Ab-initio Simulation Package (VASP) code~\cite{KRESSE199615} with the projected augmented-wave (PAW) method~\cite{PhysRevB.50.17953}, employing 13 ($3d^{10}4s^{2}4p^{1}$) and 6 ($2s^{2}2p^{4}$) valence electrons for Ga and O atoms, respectively. 
The generalized gradient approximation (GGA) with Perdew-Burke-Ernzerhof (PBE)~\cite{PhysRevLett.77.3865} functional is used for $xc$-correlation. 
The kinetic energy cutoff of the plane-wave is set to 520 eV. 
For structural optimization, the energy convergence criterion was $1 \times 10^{-5}$ eV, and the force convergence criterion was $1 \times 10^{-2}$~eV/\r A. 
Gaussian approximated smearing with a width of 0.01 eV was used. 
The MEBs of atomic migration were calculated using the CI-NEB method by VASP-vtst~\cite{jcp2000Henkelman, henkelman2000improved}, converging the forces to within $5 \times 10^{-2}$ eV/\r A. 
More details on DFT calculations can be found in the Supporting Information,  Notes 1 and 4.
The OVITO are used for the purpose of analyzing and visualizing the atomic configuration~\cite{Stukowski_2010}. 

The energy differences ($E_\mathrm{diff}$) between the TB structures and perfect step are calculated by the following formula: 
\begin{equation}\label{eq:1}
    E_\mathrm{diff} = \frac{1}{L} (E_\mathrm{TB} - E_\mathrm{perf}),
\end{equation}
where the $E_\mathrm{TB}$ and $E_\mathrm{perf}$ are the energy of the TB structures and the energy of the perfect step structure, respectively; $L$ is the length of cells in the \hkl[010] direction, which is 6.17 \r A in our work.


\providecommand{\latin}[1]{#1}
\makeatletter
\providecommand{\doi}
  {\begingroup\let\do\@makeother\dospecials
  \catcode`\{=1 \catcode`\}=2 \doi@aux}
\providecommand{\doi@aux}[1]{\endgroup\texttt{#1}}
\makeatother
\providecommand*\mcitethebibliography{\thebibliography}
\csname @ifundefined\endcsname{endmcitethebibliography}
  {\let\endmcitethebibliography\endthebibliography}{}

\end{document}